%Paper: astro-ph/9306014
%From: <malaney@chipmunk.cita.utoronto.ca>
%Date: Thu, 17 Jun 93 13:34:32 EDT

\newfam\itfam
\newfam\slfam
\newfam\bffam
\newfam\ttfam
\newfam\bmitfam
\newfam\bcalfam
\newfam\lafam

% load fonts

% load 7pt fonts

\font\sevenrm=cmr7

\font\sevenmit=cmmi7
\font\sevencal=cmsy7
\font\sevenbf=cmbx7

% load 9pt fonts

\font\ninerm=cmr9

\font\ninemit=cmmi9
\font\ninecal=cmsy9
\font\ninebf=cmbx9

% load 10pt fonts

\font\tenrm=cmr10

% load 12pt fonts

\font\twelverm=cmr10 scaled \magstep1
\font\twelvemit=cmmi10 scaled \magstep1
\font\twelvebmit=cmmi10 scaled \magstep1
\font\twelvecal=cmsy10 scaled \magstep1
\font\twelvebcal=cmbsy10 scaled \magstep1
\font\twelveex=cmex10 scaled \magstep1
\font\twelveit=cmti10 scaled \magstep1
\font\twelvesl=cmsl10 scaled \magstep1
\font\twelvebf=cmbx10 scaled \magstep1
\font\twelvett=cmtt10 scaled \magstep1
\font\twelvelasy=lasy10 scaled \magstep1

\message{main font groups,}

\catcode`@=11
\newdimen\z@
\z@=0pt
\newskip\ttglue

% \twelvepoint sets up all fonts for printing manuscripts at twelvept

\def\twelvepoint{\def\rm{\fam0\twelverm}%

  \abovedisplayskip 14pt plus 3.6pt minus 10.8pt
  \belowdisplayskip 14pt plus 3.6pt minus 10.8pt
  \abovedisplayshortskip 0pt plus 3.6pt
  \belowdisplayshortskip 8.4pt plus 3.6pt minus 4.8pt

\textfont0=\twelverm \scriptfont0=\ninerm \scriptscriptfont0=\sevenrm%
\textfont1=\twelvemit \scriptfont1=\ninemit \scriptscriptfont1=\sevenmit%
\textfont2=\twelvecal \scriptfont2=\ninecal \scriptscriptfont2=\sevencal%
\textfont3=\twelveex \scriptfont3=\twelveex \scriptscriptfont3=\twelveex%

  \def\mit{\fam1\twelvemit}
  \def\cal{\fam2\twelvecal}
  \def\ex{\fam3\twelveex}

  \textfont\bmitfam=\twelvebmit\scriptfont\bmitfam=\ninemit

  \def\bmit{\fam\bmitfam\twelvebmit}%
  \textfont\bcalfam=\twelvebcal  \def\bcal{\fam\bcalfam\twelvebcal}%
  \textfont\itfam=\twelveit \def\it{\fam\itfam\twelveit}%
  \textfont\slfam=\twelvesl \def\sl{\fam\slfam\twelvesl}%
  \textfont\bffam=\twelvebf \scriptfont\bffam=\ninebf
   \scriptscriptfont\bffam=\sevenbf \def\bf{\fam\bffam\twelvebf}%
  \textfont\ttfam=\twelvett \def\tt{\fam\ttfam\twelvett}%

  \tt \ttglue=.5em plus.25em minus.15em
 \textfont\lafam\twelvelasy
\def\la{\lafam\twelvelasy}
  \normalbaselineskip=26pt
  \setbox\strutbox=\hbox{\vrule height10.5pt depth5.0pt width\z@}
  \let\sc=\tenrm  \normalbaselines

  \rm}

%%%%%%%%%%%%%%%%%%%%%%%%%%%%%%%%%%%%%%

\hoffset=0 true in
\hsize=6.2 true in
%\hsize=6.75 true in
\voffset=0 true in
\vsize=8 true in
%\vsize=9 true in
\overfullrule=0pt
\abovedisplayskip=15pt
\belowdisplayskip=15pt
\abovedisplayshortskip=10pt
\belowdisplayshortskip=10pt

\def\gapp{\mathrel{\raise.3ex\hbox{$>$}\mkern-14mu
              \lower0.6ex\hbox{$\sim$}}}
\def\hbar{{\mathchar'26\kern-.5em{\it h}}}

% This puts pagenumbers at top of page

% This is for reference numbers

%\def\tindent#1{\indent\llap{#1}\ignorespaces}
%\def\refn{\par\hang\tindent}

%gives greater than or approx
\def\gtorder{\mathrel{\raise.3ex\hbox{$>$}\mkern-14mu
             \lower0.6ex\hbox{$\sim$}}}
%give less than or approx
\def\ltorder{\mathrel{\raise.3ex\hbox{$<$}\mkern-14mu
             \lower0.6ex\hbox{$\sim$}}}

\twelvepoint
\baselineskip=14truept plus 1pt minus 2pt
$  $
\hfill CITA/93/22
\vskip 0.5truein
\centerline {\bf Neutrino Balls and Gamma-Ray Bursts}
\medskip\medskip\medskip
\noindent
\centerline{B. HOLDOM$^1$ and R. A. MALANEY$^{2}$}

\medskip
\centerline{$^1$
Department of Physics, University of Toronto,
}
\centerline{Toronto, ON, Canada M5S 1A7.}

\medskip
\centerline{$^2$
Canadian Institute for Theoretical Astrophysics, University of Toronto,
}
\centerline{Toronto, ON, Canada M5S 1A7.}

\vskip 0.5truein \baselineskip=24truept plus 1pt minus 2pt

\centerline{ABSTRACT}

We propose a mechanism by which the neutrino emission from a supernova-type
explosion can be converted into a gamma-ray burst of total energy $\sim
10^{50}$ ergs.  This occurs naturally if the explosion is situated inside a
ball of trapped neutrinos, which in turn may lie at a galactic core.  There
are possible unique signatures of this scenario.

\medskip \centerline{1. INTRODUCTION}

The BATSE experiment on board Compton-Gamma Ray Observatory has unequivocally
demonstrated that gamma-ray bursts (GRBs) cannot originate from a local disk
distribution (Meegan {\it et al.} 1992). Although a distribution compatible
with an extended galactic halo cannot at this time be ruled out, this recent
data has spurred renewed speculation that GRBs have a cosmological origin.
The detectability threshold of BATSE is $\sim 10^{-7}$ ergs cm$^{-2}$, and
GRBs are observed with fluences extending from this threshold up to $\sim
10^{-3}$ ergs cm$^{-2}$. A detected fluence of $10^{-5}$ ergs cm$^{-2}$ at a
distance of $1$ Gpc corresponds to an isotropic energy output of $\sim
10^{50}$ ergs. Cosmological models which purport to account for the observed
GRB characteristics require the rapid release of this amount of energy,
typically within a source region of dimensions $\ltorder 100$ km in order to
be consistent with variability timescales.

Neutrino balls (Holdom 1987) are large spherical regions of trapped
degenerate neutrino gas.  Just like other cosmological defects such as
cosmic strings and monopoles, neutrino balls are remnants of symmetry
breaking processes occurring in the very early universe. They have a lower
mass bound in the supermassive range and would thus be ideal seeds for
structure formation.  Neutrino balls in an accretion mode have been associated
with the large luminosities of quasars (Dolgov and Markin 1990,1991).

Right-handed neutrinos may be trapped in a region of space within which the
local vacuum structure conveys upon them a light Majorana mass. Outside the
trapped region the right-handed neutrinos possess a large Majorana mass. The
converse is true for left-handed neutrinos. Such a circumstance can arise in
left-right symmetric electroweak theories. Subsequent to a phase transition
in the early universe, the spontaneous breaking of parity leads to the
occurrence of domain walls separating the left and right vacua. The
neutrino mass difference on either side of the wall forbids neutrino
transport from one side of the wall to the other.  All other fermions have
the same mass on both sides. The infinite domain walls must disappear and
the closed walls quickly damp to their equilibrium (spherical) form; these
are the neutrino balls. The maximum number of neutrino balls in the universe
is constrained by the mass density of the universe, but a neutrino ball per
galaxy or more is easily accommodated.  We will say nothing further here about
the production and particle physics aspects of neutrino balls (see Holdom
1987).

Neutrino balls stabilize due to the equilibration of the internal neutrino
pressure, $\rho$,  with the pressure induced by the surface tension of the
wall, $$\rho={6 \sigma\over R_b}  \ \ \ ,\eqno(1)$$ where $\sigma$ is the
surface energy density of the wall, and $R_b$ is the ball radius. The total
mass of the ball is then given by $$M_b=4\pi R_b^2 \sigma +{4\over 3} \pi
R_b^3\rho = 12\pi R_b^2 \sigma \ \ . \eqno(2)$$ If a neutrino ball is too
big it becomes a black hole, and if it is too small it rapidly disappears
via the reaction $\nu\bar\nu\rightarrow e^+e^-$. $M_b$ must lie within the
range $M_c < M_b < M_h$, where $$M_c\approx 10^4 [\sigma({\rm TeV}^3) ]^3\ \
\ M_{\odot} \ \ \  \eqno(3a)$$ $$M_h\approx 10^8/\sigma ({\rm TeV}^3) \ \ \
M_{\odot} \ \ \ . \eqno(3b)$$  This constrains ${\sigma }^{1/3}$ to be less
than a few TeV. Neutrino balls within the above mass range rapidly cool
through the reaction $\nu\bar\nu\rightarrow e^+e^-$, which then shuts off
once neutrino degeneracy is reached.  The number density and energy density
of the degenerate neutrinos are given by (assuming 3  neutrino families and
equal numbers of neutrinos and antineutrinos) $$n_\nu\approx{\mu(t)^3\over
\pi^2}  \ \eqno(4a)$$ $$\rho_\nu\approx{3\mu(t)^4\over 4\pi^2} \ \ \ .
\eqno(4b)$$ $\mu(t)<m_e$ is the chemical potential of the neutrino gas and
$m_e$ is the electron mass. The neutrino ball then very slowly shrinks and
releases energy through the reaction $\nu\bar\nu\rightarrow 3\gamma$, and
$\mu(t)$ gradually increases.  The ball lifetime in terms of the initial
$\mu_0$ is $$\tau_b\approx 10^{14} \biggl [{m_e\over \mu_0}\biggr ]^{13} \ \
\ \ {\rm secs} \ \ \ . \eqno(5)$$  The mass and radius of the ball may be
written as functions of $\mu(t)$, $$M_b(t)\approx 50 \biggl [{m_e\over
\mu(t)}\biggr ]^8 [\sigma({\rm TeV}^3) ]^3 \ \ \ M_{\odot} \ \ \eqno(6)$$
$$R_b(t)\approx 2\times 10^{10} \biggl [{m_e\over \mu(t)}\biggr ]^4
\sigma({\rm TeV}^3)  \ \ \ {\rm cm} \ \ \ . \eqno(7)$$

\medskip \centerline{2. PRIMARY SIGNAL}

If neutrino balls exist they will have a range of masses. Of most interest
for us will be balls at the high end of their mass range; that is, $M_b\sim
10^{8-9} M_{\odot}$ for $\sigma\sim 1$ TeV$^3$.  This implies $\mu\sim 50$
keV and $R_b\sim 10^{14}$ cm.  Since the chemical potential evolves so
slowly we may treat these quantities as constants.  Gravity will influence
the properties of the largest balls (Ma\'nka, Bednarek, and Karczewska
1993), but we will ignore these effects.

Let us assume that the  ball is located in the vicinity of other stars,
perhaps in a galactic core. The neutrino ball wall is transparent to all
matter except neutrinos and thus stars pass freely into the ball interior.
Stars will be captured by the gravitational field of the ball, and some will
remain in the ball interior. We will consider the sequence of events if a
supernova-type explosion were to take place within the ball. The actual type
of explosion is not important, all that we require is that a large neutrino
flux is emitted from the surface of some compact object. Since the compact
source is in a region with a parity flipped version of our vacuum, the weak
interactions are parity flipped, and therefore the neutrinos emitted from
the compact source are all right handed. We wish to determine the gamma-ray
signal caused by the interactions of these emitted neutrinos, $\nu$, with
the ambient right-handed neutrinos, $\nu'$, in the
ball.\footnote{$^\dagger$}{By ``neutrinos'' we often mean both
right-handed neutrinos and left-handed
\break
antineutrinos.}

The cross section for $\nu\bar\nu'\rightarrow e^+e^-$ is given by
$$\sigma_\nu={G_F^2\over 3\pi}\epsilon\epsilon'(1-\cos\theta)
(1\pm4\sin^2\theta_W+8\sin^4\theta_W) \ \ \ , \eqno(8)$$ where $G_F$ is
Fermi's constant,  $\epsilon$ and $\epsilon'$ are the emitted and ambient
neutrino energies, $\theta$ is the angle between their momenta, and
sin$^2\theta_W=0.23$. For the ambient neutrinos $\epsilon' \approx {3\over
4}\mu$. The  positive (negative) sign  in eq.~(8) is to be used for
$\nu_e\bar\nu_e$  ($\nu_\mu\bar\nu_\mu, \nu_\tau\bar\nu_\tau)$ interactions,
since $\nu_e\bar\nu_e$ interactions involve both charged  and neutral
currents. From eqs.~(4a) and (8) the mean free path of the electron
neutrinos is $$\lambda\approx {8\times 10^{14}\over \epsilon ({\rm MeV})}
\biggl [{m_e\over \mu} \biggr ]^4  \ \ \ {\rm cm} \ \ \ . \ \ \ \eqno(9) $$
For muon and tau neutrinos the mean free path is $\sim 5$ times larger.  We
use $\epsilon\sim 15$ MeV (30 MeV) as the average energy for electron (muon
and tau) neutrinos.  Combined with eq.~(7) this determines the number,
$N_e$, of electrons and positrons produced by the emitted electron neutrinos
during their propagation to the  neutrino-ball wall.  If this distance is of
order the radius of the ball then $${N_e \over N_\nu}\approx 8\times 10^{-4}
\sigma ({\rm TeV}^3) \ \ \ , \eqno(10) $$ where $N_\nu$ is the total number
of emitted  electron neutrinos.  The corresponding ratio
for muon and tau neutrinos is about 2/5 as large.

Each electron and positron in a produced pair will have energies close to
half of the neutrino energy and with momenta close to the outward radial
direction, since the center of mass energy of most of the pairs is not much
greater than $2m_e$.  Most will stay contained within the moving shell of
thickness $\Delta \tau_\nu  c$ where $\Delta \tau_\nu$ is the duration of
the burst.  For a typical supernova $\Delta \tau_\nu \sim 10$ secs.  Only
when the radius of the shell approaches $10^{14}$ cm $\sim R_b$ will the
electron shell start to significantly spread in the radial direction, due to
velocities which are not exactly $c$.

We now consider the self-interactions of the $e^+e^-$ pairs.  We will have to
require that they do not interact so much that they thermalize and
significantly reduce their average energy.  Thermalization can only happen if
the total number of electrons, positrons, and photons in the shell increases.
The cross sections for number changing interactions, such as bremsstrahlung,
contain an extra factor of $\alpha$ and will be approximated by
$\alpha\sigma_T$ where $\sigma_T$ is the Thomson cross section.  With this we
estimate the total increase in the number of particles including photons in
the shell compared to the total number of electrons and positrons produced
from neutrino scattering:  $${\Delta N \over N_e}=\int_{0}^{R_b/c}n_e(t)
\alpha \sigma_T\Delta v{N_e(t) \over N_e}dt \ \ \ . \eqno(11)$$  $\Delta
v$ is a typical {\it relative} velocity of electrons and
positrons.  We may write $N_e\approx 2{R_b \over \lambda}{{E}_{i} \over
{\epsilon }}$, $N_e(t)={ct \over R_b}N_e$, and $n_e(t)=N_e(t)/V(t)$, where
$E_i$ is the total initial energy in neutrinos and $V(t)$ is the volume of the
shell.  The result is $${\Delta N \over N_e}\approx 10^6 {E_i
(10^{53}{\rm ergs})\over \Delta \tau_\nu ({\rm secs}) }{\Delta v \over c}
{\left({\mu  \over {m}_{e}}\right)}^{4} \ \ \ . \eqno(12) $$  With $\Delta
v \sim c/10$, $E_i\sim 10^{53}$ ergs, and $\mu =50$ KeV we find ${\Delta N
\over {N}_{e}}\approx 1$.

We have thus found that $\mu $ cannot be much greater than 50 KeV in order
for the spectrum of gamma-rays to remain largely in the 1-10 MeV energy
range.  When $\mu$ is of this order then we are also guaranteed
that the total energy is equi-partitioned among the electrons, positrons and
photons, due to the lowest order electromagnetic interactions. For a typical
supernova event in which $2 \times 10^{53}$ ($4 \times 10^{53}$) ergs is
carried away by emitted electron (muon and tau) neutrinos, eq.~(10) implies
$\sim 10^{50}$ ergs end up in the gamma-ray burst, the duration of which is
$\Delta \tau_\nu$. This then is the expected signature of the initial
propagation of neutrinos emitted by a supernova within a neutrino ball of
sufficient size.  For neutrino balls much smaller than $R_b\approx 10^{14}$
cm, for which $\mu$ is significantly larger, the signal is likely damped.

We now comment on how the ambient neutrino gas can affect
stellar evolution, as a consequence of scattering with the stellar matter.
We first note that the chemical potential of the degenerate neutrino gas is
larger than typical stellar temperatures.  But precisely because of the
degeneracy, the collisions which could serve to heat the stellar matter do
not take place due to phase-space blocking.  Instead, an ambient neutrino must
absorb sufficient energy to emerge above the Fermi surface.  We find that
this can constitute a stellar cooling mechanism several orders of magnitude
larger than ordinary neutrino losses. An exact description of how this
alters the stellar evolution would require detailed stellar modeling.
However, a rapid increase in the evolutionary timescales could be expected,
and this would enhance the probability of supernova explosions
occurring within the neutrino ball.

The ambient neutrino gas also produces a dynamical friction for stellar
motion.  The timescale for  the star to  settle at the center of the ball
through dynamical friction will be  $$\tau_{f}\sim {M_b\over m_s}
{1\over\sqrt {G\rho}} \ \ \ , \eqno(13)$$ where $G$ is Newton's constant.
For our adopted ball parameters and assuming $m_s=10M_{\odot}$, we find
$\tau_f\sim 10^4$ years. The star may therefore settle at the ball's center
in a relatively short timescale. This in turn raises the possibility of
stellar mergers at the ball's center leading to the formation of massive
stars, or to enhanced collisions of white dwarfs and neutron stars. These
effects further increase the likelihood of supernova explosions within the
ball.\footnote{$^\dagger$}{A detailed discussion of mergers in dense stellar
environments can be found in Quinlan and Shapiro (1990).}

\vfill\eject
\medskip \centerline{3. SECONDARY SIGNALS FROM REFLECTED NEUTRINOS}

Additional distinctive signatures of our picture could occur due to the
peculiar reflective property of a neutrino-ball wall. On leaving the
supernova the neutrinos expand with only a small fraction converted to the
primary $e^+e^-\gamma$ shell. However, on reaching the wall the neutrinos are
energetically forbidden from entering the other vacuum, and are therefore
reflected back into the interior of the ball.  As the neutrino shell
propagates back in from the wall it generates another $e^+e^-\gamma$ shell
traveling with it.  The fate of this secondary $e^+e^-\gamma$ shell can be
influenced by any $e^+e^-$ gas left behind  the primary $e^+e^-\gamma$
shell.  The source of particles lagging behind the primary $e^+e^-$ shell
could be due to the bremsstrahlung type interactions considered above, or
due to the initial shell interacting with any ambient particles already in the
ball. The scattering of secondary-shell particles   with the
 remnant $e^+e^-$ gas
no longer involves the   factors of $\alpha$ or $\Delta v/c$ explicitly shown
in eq.~(11). Such scatterings can degrade the energy of the particles.
If a significant fraction of the
electrons and positrons produced in the
primary burst are left behind in the ball ($\sim$ a few percent) then
substantial
damping of the  secondary gamma-ray signal could occur.

We will proceed in this section by assuming that this damping of the
secondary signal does not happen, at least in some cases.  Even then the
secondary shells of $e^+e^-$ pairs will still collide with other $e^+e^-$
shells traveling in other directions.
One could easily envisage a large variety of signals depending on the
position of the  explosion within the ball.  For explosions sites not
on center one could expect that after several reflections the neutrinos
propagate no longer in thin shells, but instead propagate more randomly
throughout the ball. $e^+e^-$ pairs and photons are then produced uniformly
over the ball.  The expected signal for this process would possibly include
some secondary gamma-ray bursts from the first few reflections.  But the
signal would gradually turn into a continuous gamma-ray flux
$\Delta t/R_b$ ($\sim 10^{-3}$ ) times smaller than the original
burst,  lasting for a total time $\lambda/c$ ($\sim 1$ month).

In the special case where the explosion event occurs near the
ball's center an interesting phenomena can take place -- after reflection
the emitted neutrinos  will be {\it re-focussed} at the center of the ball.
That the explosion event may take place at the ball's center does not seem
too unreasonable considering our discussion of eq.~(13).  Assuming  an
on-center explosion,  we wish to determine what fraction of the neutrino
energy will be lost to $\nu\bar\nu \rightarrow e^+e^-$ pair production  when
the  neutrinos re-focus at the center.  If we consider a re-focus region of
the same size $r_\nu$ as the original neutrinosphere of the supernova then
we find that the ratio of energy lost to pair production to the initial
energy of explosion is approximated by $${E_p\over E_i}\sim
10^{-3}{E_i(10^{53}{\rm ergs}) \epsilon({\rm MeV})\over r_\nu(10^6{\rm cm})
\Delta \tau_\nu ({\rm secs}) } \ \ \ . \eqno(14)$$ For a typical explosion
event then, roughly $10^{-3}E_i$ is removed from the neutrino energy during
the re-focusing.

Since energy losses are small this means that after the neutrinos pass
through the  re-focusing region most of them repeat their cycle. That is,
they  propagate  isotropically to the ball wall from which they are once
again are reflected back toward the center. The types of signal we have just
calculated in the previous section can therefore be repeated with a cycle
time of $2R_b/c$.  The  duration
of this signal  -- determined by damping effects as discussed above --
can be  no larger than $\lambda/c$.
Nonetheless, we wish to emphasize
 the possibility of a unique {\it periodic}
signal with  decreasing amplitude.

In addition to signals produced through interaction with the ambient
neutrinos, there remains the possibility of other gamma-ray signals -- as a
consequence of $\nu\bar\nu$-induced fireballs. The physics of fireballs has
been developed in a series of previous papers (eg. Cavallo and Rees 1978;
Goodman 1986; Paczy\'nski 1986, 1990; Shemi and Piran 1990; Piran and Shemi
1993). Assuming thermal equilibrium, the initial temperature $T_i$ of a
fireball  is determined only by  the rate of energy input, $\dot E$, and
the  radius $R_i$ within which the initial energy is confined. For $\dot
E\sim 10^{50}$ ergs s$^{-1}$ and $R_i\sim 10^6$cm, we find $T_i\sim 1$ MeV.
In the outer regions of the fireball a relativistic outflow is formed, and
the apparent temperature of the escaping photons is approximately equal to
$T_i$.

Using eq.~(14) we have found that $\sim 10^{50}$ ergs can be input into the
fireball. Re-focusing of the $e^+e^-$ pairs produced by the incoming
neutrinos gives a similar energy input.  However, for on-center explosions
it is unlikely that any gamma-ray signal will be observed. The main reason
for this is the remnant supernova debris around the explosion site which is
opaque to photons. There is also the effect of baryons on the fireball
itself, since the thermal energy of the fireball tends to be diverted to
kinetic energy of the baryons.  Fireball effects, however, are not
completely ruled out. For example, suppose that the source is some distance
$r$ off-center with $ r << R_b$. Then the emitted neutrinos are re-focussed
in a region diametrically opposite to the explosion site. For
some range of $r$ it is possible that some re-focusing occurs in regions
where the remnant debris has yet to reach, and observable fireballs are
produced. There also remains the possibility of break-out and beaming
effects -- similar to that discussed by M\'ez\'aros and Rees (1993) for
colliding neutron stars -- arising from anisotropic debris distributions.

\medskip \centerline{4. DISCUSSION}

We have seen that supernova-type explosions within the confines of neutrino
balls can produce the gross characteristics required for the production of a
cosmological gamma-ray burst (GRB). The interaction of emitted neutrinos with
ambient neutrinos can readily lead to a $10^{50}$ erg GRB with a
characteristic timescale of $10$ secs.  The latter timescale is set by the
supernova itself.  In the case that the ball remains transparent to
photons after the initial pulse, then unique observational signatures can
possibly arise from the neutrinos reflected back into the ball. For
off-center events we could have one or more weaker secondary pulses before
the signal fades into a continuous ``afterglow'' of duration $\sim
\lambda/c$.  An on-center event could produce a strictly periodic GRB with
decreasing amplitude, and period $2R_b/c$.

We have mentioned various ways in which supernova-type events may be
induced inside neutrino balls, including the cooling and rapid evolution of
stars, the merging of stars into more massive stars, and collisions
involving white dwarfs or neutron stars.  All these possibilities -- coupled
with our lack of knowledge on the number and size of neutrino balls in the
universe --  make it difficult to predict the frequency of the events we
have described here.

Neutrino balls may be associated with other exotic phenomena; colliding
neutrino balls and the death of neutrino balls are but two examples. In the
latter case the slow evolution of the neutrino ball comes to an abrupt end
when the chemical potential $\mu$ becomes of order the electron mass. At
this point the reaction $\nu\bar\nu\rightarrow e^+e^-$ for ambient
neutrinos, which previously was energetically forbidden, now proceeds
rapidly. We find that the time scale for the conversion of the neutrinos
into $e^+e^-$ pairs is a few hours.

It may well be that the strange and perplexing phenomena of GRBs are
indicating the presence in the universe of some new exotic configuration. If
so, neutrino balls may be an ideal candidate. The most obvious test for the
existence of a  cosmological neutrino ball --
optical observation of an extragalactic
Type~II supernova with no apparent neutrino burst -- is not technically
feasible. Continued scrutiny of GRB signals for the various secondary
signals we have discussed would seem to be the best alternative.

\medskip\medskip\medskip
\noindent
RAM acknowledges useful discussions with  his CITA colleagues.  This work
was supported in part by Natural Sciences and Engineering Research Council of
Canada.

\medskip\medskip

\centerline{REFERENCES}

\item{}Cavallo, G. and Rees, M. J., 1992 {\it M.N.R.A.S.}, {\bf 183}, 359.

\item{}Dolgov, A.D. and Markin, O. Yu., 1990, {\it Sov. Phys. JETP}, {\bf
71}, 207.

\item{}Dolgov, A.D. and Markin, O. Yu., 1991, {\it Prog. Theor. Phys.},
{\bf 85}, 1091.

\item{}Goodman, J., 1986, {\it Ap. J. Letts.}, {\bf 308}, L47.

\item{}Holdom, B, 1987, {\it Phys. Rev. D}, {\bf 36}, 1000.

\item{}Ma\'nka, R., Bednarek, I. and Karczewska, 1993, preprint
(U\'SL-TH-93-01),
astro-ph/9304007.

\item{}Meegan, C. A., Fishman, G. J., Wilson, R. B., Paciesas, W. S.,
Pendleton, G. N.,
\ \ \ Horack, J. M., Brock, M. N. and Kouveliotou, C., 1992, {\it Nature}, {\bf
355}, 143.

\item{}M\'esz\'aros, P. and Rees, M. J., 1992 {\it M.N.R.A.S.}, {\bf 257}, 29p.

\item{}Paczy\'nki, B., 1986, {\it Ap. J. Letts.}, {\bf 308}, L43.

\item{}Paczy\'nki, B., 1990, {\it Ap. J. Letts.}, {\bf 363}, 218.

\item{}Quinlan, G. D. and Shapiro, S. L., 1990, {\it Ap. J.}, {\bf 356}, 483.

\item{}Piran, T. and Shemi, A., 1993, {\it Ap. J. Letts.}, {\bf 403}, L67.

\item{}Shemi, A. and Piran, T. , 1990, {\it Ap. J. Letts.}, {\bf 365}, L55.

\bye